\documentclass[aps,pra,floatfix,superscriptaddress,twocolumn]{revtex4}
\usepackage[dvips]{graphicx}
\usepackage{color}

\usepackage{longtable}
\usepackage{dcolumn}
\usepackage[dvips]{graphicx}
\usepackage{bm}
\usepackage{bbm}

\usepackage{times}
\usepackage{nicefrac}
\usepackage{amsmath}
\usepackage{amsfonts}
\usepackage{amssymb}
\usepackage{amsthm}

\newcolumntype{.}{D{x}{}{-1}}

\newcommand{\balpha}{\bm{\alpha}}

\newcommand{\bnabla}{\bm{\nabla}}

\newcommand{\vare}{\varepsilon}
\newcommand{\bfr}{{\bm {r}}}

\newcommand{\bfp}{{\bm {p}}}

\newcommand{\hr}{\hat{\bfr}}

\newcommand{\lbr}{\langle}
\newcommand{\rbr}{\rangle}

\newcommand{\Za}{Z\alpha}

\begin{document}

\title{Energy levels of core-excited $\bm{1s2l2l'}$ states in lithium-like ions: argon to uranium}

\author{V.~A. Yerokhin}
\affiliation{Physikalisch-Technische Bundesanstalt, D-38116 Braunschweig, Germany}
\affiliation{Center for Advanced Studies, Peter the Great St.~Petersburg Polytechnic University,
195251 St.~Petersburg, Russia}

\author{A.~Surzhykov}
\affiliation{Physikalisch-Technische Bundesanstalt, D-38116 Braunschweig, Germany}
\affiliation{Technische Universit\"at Braunschweig, D-38106 Braunschweig, Germany}

\begin{abstract}

Energy levels and fine-structure intervals of the $1s2l2l'$ core-excited states are calculated
for ions along the Li isoelectronic sequence from argon to uranium. The calculation is performed
by the relativistic configuration-interaction method adapted for treatment of autoionizing
core-excited states. The calculational approach includes the relativistic treatment of the
nuclear recoil effect, the leading QED shifts as delivered by the model QED operator, and the
frequency dependence of the Breit interaction. The $1s2l2l'$--$1s^22l$ transition energies are
obtained by combining the present results for the $1s2l2l'$ states with energies of the $1s^22l$
states compiled from previous calculations. All theoretical energies are supplied with
uncertainty estimates. Our theoretical predictions for the $1s2l2l'$--$1s^22l$ transitions are
significantly more accurate than the best experimental results available today and can be used
for calibrating experimental X-ray spectra.

\end{abstract}

\maketitle

\section{Introduction}

Spectroscopic data on dielectronic satellite spectra of highly charged ions are widely used for
analysing stellar flares, tokamak and laser-produced plasmas. Such spectra provide important
information on the electron temperature and density, the ionization state distribution, and other
characteristics of hot plasmas, which are required in non-perturbing spectroscopic techniques for
plasma diagnosis.

There are extensive tabulations of atomic spectroscopic data available in the literature, which are
widely used for modelling atomic spectra and astrophysical analysis, {\em e.g.},
Refs.~\cite{shirai:00,sansonetti:05,nist}. These tabulations are mostly based on critically
evaluated experimental results and include theoretical values for H-like and sometimes for He-like
ions. The reason for this is that theoretical calculations from 1980s and 1990s for ions with a
larger number of electrons were often not very reliable and could not provide quantitative
predictions with estimations of uncertainty of uncalculated effects.

Significant progress in theory of highly charged ions has been achieved in the last years. For
Li-like ions, rigorous QED calculations to all orders in the nuclear strength parameter $\Za$
(where $Z$ is the nuclear charge number and $\alpha$ is the fine structure constant) were performed
for the ground and the first valence excited states
\cite{yerokhin:01:2ph,sapirstein:01:lamb,yerokhin:06:prl,yerokhin:07:lilike,kozhedub:10,sapirstein:11}.
Extensive calculations of energies of core-excited states of Li-like ions were carried out in our
previous investigations \cite{yerokhin:12:lilike,yerokhin:17:lilike}. All these calculations
yielded {\em predictive} results, {\em i.e.}, results with estimations of uncertainties due to
uncalculated effects. Moreover, for the core-excited states, the theoretical predictions were shown
to be significantly more accurate than the best experimental energies available today (see the
comparison in Tables VII-IX of Ref.~\cite{yerokhin:17:lilike}). This conclusion is confirmed in the
present work by comparing theoretical values against benchmark experimental results in the medium-
and high-$Z$ region \cite{gonzales:05:prl,gonzales:06,rudolph:13}.

High accuracy of theoretical energies of core-excited states of Li-like ions makes them the
preferable source of spectroscopic data for modelling plasma spectra and opens up new possibilities
of their usage for calibration of experimental X-ray spectra for ions with a larger number of
electrons \cite{yerokhin:17:lilike}.

The goal of the present work is to extend our previous calculation of energies of core-excited
$1s2l2l'$ states of Li-like ions with $Z = 6$ -- $17$ in Ref.~\cite{yerokhin:17:lilike} to the
higher-$Z$ region, $Z = 18$ -- $92$, thus applying the same method across the entire isoelectronic
sequence, and evaluating theoretical uncertainties in the same way for all nuclear charges.

Relativistic units $\hbar = c = 1$ and $\alpha = e^2/(4\pi)$ are used throughout this paper.

\section{Method of calculation}

\subsection{Dirac-Coulomb-Breit energies}

The no-pair Dirac-Coulomb Breit (DCB) Hamiltonian is given by
\begin{eqnarray} \label{eq:DCB}
    H_{\rm DCB} = \sum_i h_{\rm D}(i) + \Lambda_{++} \sum_{i<j} \left[ V_{C}(i,j)+
    V_{B}(i,j)\right]\,\Lambda_{++} \,,
\end{eqnarray}
where the indices $i,j = 1,\ldots,N$ numerate the electrons, $h_D$ is the one-particle Dirac
Hamiltonian with the nuclear Coulomb potential $V_{\rm nuc}$ and the Dirac-Fock potential of the
frozen core $V^{N-1}_{\rm DF}$,
\begin{equation}\label{eq5a}
    h_{D} = \balpha\cdot\bfp+ (\beta-1)\,m+ V_{\rm nuc}(r)+
    V^{N-1}_{\rm DF}(\bfr)\,,
\end{equation}
$\Lambda_{++}$ is the projector on the positive-energy states of the Hamiltonian $h_D$, and $V_{C}$
and $V_B$ are the Coulomb and the Breit parts of the electron-electron interaction,
\begin{align}\label{eq3}
V_C(i,j) &\ = \frac{\alpha}{r_{ij}}\,,\\
 \label{eq3a}
V_{B}(i,j) &\ =  -\frac{\alpha}{2\,r_{ij}}\,
    \left[ \balpha_i\cdot\balpha_j + \left( \balpha_i\cdot \hr_{ij}\right)
             \left( \balpha_j\cdot \hr_{ij}\right) \right]\,,
\end{align}
$\balpha = \gamma^0{\bm\gamma}$ and $\beta = \gamma^0$ are the Dirac matrices, $\bfr_{ij} =
\bfr_i-\bfr_j$,  and $\hr = \bfr/r$.

The eigenvalues of the DCB Hamiltonian are obtained by the configuration-interaction (CI) method.
In this method, the $N$-electron wave function $\Psi(PJM)$ with a definite parity $P$, total
angular momentum $J$, and angular momentum projection $M$ is represented as a finite sum of
configuration-state functions (CSFs) with the same $P$, $J$, and $M$,
\begin{equation}\label{eq4}
  \Psi(PJM) = \sum_r c_r \Phi(\gamma_r PJM)\,,
\end{equation}
where $\gamma_r$ denotes the set of additional quantum numbers that determine the CSF. The CSFs are
constructed as linear combinations of antisymmetrized products of one-electron orbitals, which are
positive-energy eigenfunctions of $h_D$. Energy levels of the atom and the wave-function expansion
coefficients $c_r$ are obtained as the eigenvalues and the eigenvectors of the matrix of the DCB
Hamiltonian in the space of the CSFs,
\begin{eqnarray}\label{eq:0}
   \left\{ H_{rs}\right\} \equiv \left\{ \lbr \gamma_r PJM|H_{\rm DCB}|\gamma_s PJM\rbr\right\}\,.
\end{eqnarray}
Details of our implementation of the CI method can be found in
Refs.~\cite{yerokhin:08:pra,yerokhin:12:lilike,yerokhin:17:lilike}.

The energy levels obtained from the DCB Hamiltonian (\ref{eq:DCB}) are supplemented by various
corrections considered below.

\subsection{Frequency-dependent Breit interaction}

The operator of the electron-electron interaction in the form of $V_C+V_B$ [as given by
Eqs.~(\ref{eq3}) and (\ref{eq3a})] is obtained within the Breit approximation. It does not depend
on the energy (frequency) of the virtual exchange photon and thus neglects the retardation effects.
The exact QED operator of the electron-electron interaction has an explicit dependence on the
virtual photon energy $\omega$. The frequency-dependence of the electron interaction gives rise to
the following additional term that should be added to the Coulomb and the unretarded Breit
interactions,
\begin{align}\label{eq:vbret}
V_{B,\rm ret}&\ (\omega, i,j) = I_{\rm Coul}(\omega,i,j) - V_C(i,j) - V_B(i,j)
 \nonumber \\
&\ = \alpha\,\Biggl[\balpha_i\cdot\balpha_j\,\frac{1-\cos(\omega r_{ij})}{r_{ij}}
 \nonumber \\ &
 + \bigl( \balpha_i\cdot\bnabla_i\bigr)\, \bigl( \balpha_j\cdot\bnabla_j\bigr) \,
  \frac{\cos(\omega r_{ij})-1+ \omega^2 r^2_{ij}/2}{\omega^2 r_{ij}}\Biggr]\,,
\end{align}
where $I_{\rm Coul}(\omega,i,j) = \alpha \,
\alpha_{\mu}\alpha_{\nu}D^{\mu\nu}(\omega,|\bfr_i-\bfr_j|)$ is the (real part of the) QED operator
of the electron-electron interaction in the Coulomb gauge and $D^{\mu\nu}$ is the photon propagator
(see Ref.~\cite{shabaev:02:rep}).

In our calculation, we include the frequency-dependent part of the electron-electron interaction
{\em only} for the reference-state configuration(s), which corresponds to the correct treatment of
the one-photon exchange correction in the QED perturbation theory \cite{shabaev:02:rep}. We note
that accounting for the frequency dependence beyond that level is possible only within a systematic
QED approach (see Refs.~\cite{yerokhin:00:prl,yerokhin:07:lilike} for details). Because of this,
inclusion of the frequency-dependent interaction into all elements of the DBC Hamiltonian matrix
(\ref{eq:0}) can yield spurious effects and should be avoided.

\subsection{Relativistic recoil}

Since in the present work we are interested in medium- and high-$Z$ ions, we need to treat the
nuclear recoil effect relativistically, beyond the standard nonrelativistic normal and specific
mass-shift operators. The relativistic recoil operator was derived in
Refs.~\cite{shabaev:85,shabaev:88} (see also Ref.~\cite{shabaev:02:rep}),
\begin{eqnarray}\label{eq:rrec}
H_{\rm rec} = \frac{m}{2M}\, \sum_{ij}\biggl[ \bfp_i\cdot\bfp_j -\frac{Z\alpha}{r_i} \left( \balpha_i\cdot\bfp_j
+ \frac{\balpha_i\cdot\bfp_i\,\bfr_i\cdot\bfp_j}{r_i^2}\right) \biggr]\,,
\end{eqnarray}
where $m$ and $M$ are the mass of the electron and the nucleus, respectively. In order to compute
the energy shifts due the relativistic recoil effect, we perform our CI calculations with and
without the operator $H_{\rm rec}$ added to the DCB Hamiltonian. Since the resulting energy shift
is relatively small, it is sufficient to use a smaller basis of CSFs than in the main part of the
CI calculation.

\subsection{Leading QED effects}
\label{sec:qed}

The QED effects must nowadays be included in any accurate calculations of atomic energy levels.
Rigorous QED calculations have been performed for the $1s^22s$ and $1s^22p_j$ states of Li-like
ions, but not for other states. In our present treatment of core-excited states, we utilize the
method of the model QED operator developed by Shabaev {\em et al.} \cite{shabaev:13:qedmod}. In
this method, the exact one-loop QED operator is replaced by the approximate operator $V_{\rm QED}$,
\begin{align}\label{eq:qedmod}
V_{\rm QED} = V_{\rm SE, loc}(r) + V_{\rm SE, nloc} + V_{\rm Uehl}(r) + V_{\rm WK}(r)\,,
\end{align}
where $V_{\rm SE, loc}(r)$ is the local part of the electron self-energy, $V_{\rm SE, nloc}$ is the
non-local remainder of the self-energy operator, $V_{\rm Uehl}(r)$ and $V_{\rm WK}(r)$ are the
Uehling and the Wichmann-Kroll parts of the vacuum polarization \cite{mohr:00:rmp}, respectively.
The nonlocal part of the self-energy operator is defined as \cite{shabaev:13:qedmod}
\begin{align}\label{eq:qedmod2}
V_{\rm SE, nloc} = & \sum_{ijlk = 1}^n |\phi_i \rbr \,
\bigl( D^{{-1}}\bigr)_{ji}\,
  \nonumber \\ \times &
\biggl< \psi_j \biggl| \biggl[\frac12 \Sigma(\vare_j)+ \frac12 \Sigma(\vare_l) - V_{\rm SE, loc}\biggr] \biggr| \psi_l\biggr>\,
  \nonumber \\ \times &
\bigl( D^{-1}\bigr)_{lk}\,
\lbr \phi_k|\,,
\end{align}
where $\Sigma(\vare)$ is the exact one-loop self-energy operator \cite{yerokhin:99:pra}, $\psi_i$
are the hydrogenic wave functions, $\phi_i$ are the model wave functions, and $D$ is the overlap
matrix, $D_{ij} = \lbr \phi_i|\psi_j\rbr$. For further details on the construction of the model QED
operator we refer the reader to the original work \cite{shabaev:13:qedmod}.

In this study we use the model QED operator as implemented by the QEDMOD package
\cite{shabaev:14:qedmod,shabaev:18:qedmod}. Specifically, the model QED potential is added to the
one-particle Hamiltonian $h_D$ in Eq.~(\ref{eq:DCB}), $h_D \to h_D + V_{\rm QED}$, for one-electron
matrix elements between orbitals $i$ and $j$ if {\em both} of them are from the {\em discrete} part
of the spectrum (i.e., $-mc^2 < \vare_{i,j} < 0$).

The uncertainty induced by the approximate nature of the model QED operator was estimated by
comparing the QED shifts for the $1s^22s$ and $1s^22p_j$ states as obtained with the QEDMOD package
with those from the rigorous QED treatment. Based on the detailed analysis presented in our
previous study \cite{yerokhin:17:lilike} (see Table II therein) and extended in the present work,
we estimate the QED uncertainty of the total binding energy of a state as $(18/Z)\%$ of the QED
correction (for the region of $Z\ge 18$ addressed in the present work). This is consistent with the
corresponding estimate of 1\% for $Z<18$ in in our previous work \cite{yerokhin:17:lilike}.

For calculating the QED uncertainty in transition energies, in most cases we propagate errors in
the standard way, adding quadratically the uncertainties of the two transition states. For some
transitions (particularly, for the fine-structure intervals), however, there is a large
cancellation of the QED effects in the difference. In these cases, the QED corrections for the two
transition states are highly correlated because the dominant contribution comes from the
core-electron charge density, which does not change much between the initial and the final state.
In these cases of large cancellations, we estimate the QED uncertainty as 4\% of the QED correction
to the energy difference.

\subsection{Higher-order QED effects}

Two higher-order QED effects were also included in our calculation, namely, the two-loop QED
effects and the QED part of the nuclear recoil. Both these effects were accounted for in the
approximation of independent electrons, by using data tabulated in Ref.~\cite{yerokhin:15:Hlike}.
The corresponding energy shifts for the core-excited states turned out to be quite small, but we
still included them in order to be consistent with the energies of $1s^22s$ and $1s^22p_j$ states,
deduced from rigorous QED calculations as described in the next section.

\section{Calculation}

Computational details of our implementation of the CI method are described in our previous studies
\cite{yerokhin:08:pra,yerokhin:12:lilike,yerokhin:17:lilike} and will not be repeated here. It is
important to point out, however, that since we are presently interested in the autoionizing
core-excited states, it was important to use the procedure for ``balancing'' the basis set of
one-electron orbitals, which was developed in our previous investigation \cite{yerokhin:17:lilike}.
The usage of the balanced basis allowed us to significantly improve the convergence of our CI
energies with respect to the size of the basis set.

With the help of the CI method we obtain  total binding energies of the core-excited states, which
cannot be readily accessed experimentally. In order to get transition energies typically measured
in experiments, we need in addition theoretical results for the $1s^2$ state of He-like ions and
the $1s^22s$, $1s^22p_{1/2}$, and $1s^22p_{3/2}$ states of Li-like ions. Rigorous QED calculations
are available in the literature for these states
\cite{artemyev:05:pra,yerokhin:01:2ph,sapirstein:01:lamb,yerokhin:06:prl,yerokhin:07:lilike,kozhedub:10,sapirstein:11}.
These studies were mostly focused on transition energies, so we had to combine results from
different works in order to get the total binding energies. Specifically, we added together the
ionization energy of the hydrogenic $1s$ state tabulated in Ref.~\cite{yerokhin:15:Hlike}, the
ionization energy of the $1s^2$ state calculated in Ref.~\cite{artemyev:05:pra}, and the ionization
energy of a $1s^22l$ state \cite{yerokhin:07:lilike,kozhedub:10}, thus obtaining the total binding
energy of the $1s^22l$ state, see Supplementary Material for details.

The resulting energies of the $1s^22l$ states are listed in Table~I of Supplementary Material.
Table~II of Supplementary Material presents a comparison of the compiled energies of the
$2p_{1/2}$-$2s$ and $2p_{3/2}$-$2s$ transitions with the theoretical results of
Refs.~\cite{yerokhin:07:lilike,kozhedub:10,sapirstein:11}.

As an important cross-check, we apply our present CI-QEDMOD approach to the $1s^2$, $1s^22s$, and
$1s^22p_{j}$ states and compare the resulting energies with those obtained in {\em ab-initio} QED
calculations. The corresponding comparison is presented in Table~\ref{tab:qedcompare}. We observe a
very good consistency between the two approaches and conclude that our estimation of the QED
uncertainties is adequate.

The region of the nuclear charges covered in the present work, $Z = 18$ -- $92$, overlaps with the
region $Z = 18$ -- $36$ for which an analogous calculation was carried out by us in 2012
\cite{yerokhin:12:lilike}. The differences with our previous calculation are as follows. First, we
now use the balanced basis set of one-electron orbitals (see Ref.~\cite{yerokhin:17:lilike} for
details), which allows us to improve the numerical accuracy of the DCB energies (particularly, for
the $1s2p^2\,^2\!D$ states). Second, our present approach for treatment of QED effects (with the
QEDMOD package) is completely independent from the approach used in our earlier work, and the
estimations of errors also. Our present way of estimating the uncertainty due to QED effects
(described in Sec.~\ref{sec:qed}) yields QED errors that are 2-3 times larger than those in
Ref.~\cite{yerokhin:12:lilike}, thus leading to more conservative estimates. Third, we now compile
energies of the lowest-lying  $1s^2$, $1s^22s$,  and $1s^22p_{j}$ states from published results of
{\em ab-initio} QED calculations, instead of calculating them within the same method as for the
core-excited states, as in our previous investigation. Fourth, we presently treat the nuclear
recoil effect relativistically, rather than nonrelativistically as previously.

The differences described above make our present calculation largely independent from the 2012
computation \cite{yerokhin:12:lilike}, so that a comparison between them provides an additional
cross-check. We find generally good agreement between the two calculations. In a few cases,
however, there are notable deviations: $2\,\sigma$ for $Z = 18$ and $3\,\sigma$ for $Z = 19, 20$
for the $^2\!D_J$ states; $2\,\sigma$ for $Z = 23$, the $^2\!P^o_J$ states. The main reason for
these deviations was an insufficient convergence of the CI energies in the previous calculation due
to the interaction of the reference state with continuum.

\begin{table*}
\caption{Comparison of theoretical total binding energies of the $1s^2$, $1s^22s$, and $1s^22p_{j}$
states (in Rydbergs), as obtained by the present CI-QEDMOD approach and compiled from results of rigorous QED calculations.
Uncertainties due to nuclear radii are not shown.
\label{tab:qedcompare}
}
\begin{ruledtabular}
\begin{tabular}{llddd}
\multicolumn{1}{l}{$Z$} & \multicolumn{1}{c}{State} & \multicolumn{1}{c}{CI-QEDMOD}
                          & \multicolumn{1}{c}{Full QED}          & \multicolumn{1}{c}{Difference}\\
\hline\\[-5pt]
20 & $1s^2$                                             &     -778.990\,0\,(21) &     -778.991\,66\,(2) &       -0.0016\,(21) \\
   & $1s^22s$                                           &     -864.082\,7\,(22) &     -864.084\,11\,(4) &       -0.0014\,(22) \\
   & $1s^22p_{1/2}$                                     &     -861.439\,9\,(21) &     -861.441\,02\,(4) &       -0.0011\,(21) \\
   & $1s^22p_{3/2}$                                     &     -861.067\,6\,(21) &     -861.068\,83\,(5) &       -0.0012\,(21) \\[1ex]
40 & $1s^2$                                             &     -3\,215.793\,(11) &   -3\,215.799\,95\,(34) &        -0.007\,(11) \\
   & $1s^22s$                                           &     -3\,594.090\,(11) &   -3\,594.094\,96\,(37) &        -0.005\,(11) \\
   & $1s^22p_{1/2}$                                     &     -3\,588.136\,(11) &   -3\,588.139\,47\,(36) &        -0.004\,(11) \\
   & $1s^22p_{3/2}$                                     &     -3\,580.632\,(11) &   -3\,580.636\,66\,(37) &        -0.005\,(11) \\[1ex]
60 & $1s^2$                                             &     -7\,486.216\,(27) &    -7\,486.233\,1\,(36) &        -0.017\,(27) \\
   & $1s^22s$                                           &     -8\,393.322\,(29) &    -8\,393.331\,7\,(36) &        -0.010\,(29) \\
   & $1s^22p_{1/2}$                                     &     -8\,383.086\,(27) &    -8\,383.095\,1\,(36) &        -0.009\,(27) \\
   & $1s^22p_{3/2}$                                     &     -8\,339.726\,(27) &    -8\,339.736\,2\,(36) &        -0.010\,(28) \\[1ex]
80 & $1s^2$                                             &    -13\,966.390\,(54) &    -13\,966.428\,(22) &        -0.038\,(58) \\
   & $1s^22s$                                           &    -15\,696.876\,(58) &    -15\,696.896\,(23) &        -0.019\,(62) \\
   & $1s^22p_{1/2}$                                     &    -15\,680.625\,(54) &    -15\,680.649\,(22) &        -0.024\,(58) \\
   & $1s^22p_{3/2}$                                     &    -15\,522.650\,(54) &    -15\,522.671\,(22) &        -0.021\,(59) \\[1ex]
90 & $1s^2$                                             &    -18\,252.678\,(72) &    -18\,252.762\,(54) &        -0.084\,(90) \\
   & $1s^22s$                                           &    -20\,540.196\,(79) &    -20\,540.252\,(55) &        -0.056\,(96) \\
   & $1s^22p_{1/2}$                                     &    -20\,520.271\,(73) &    -20\,520.341\,(54) &        -0.069\,(91) \\
   & $1s^22p_{3/2}$                                     &    -20\,244.332\,(73) &    -20\,244.390\,(54) &        -0.058\,(91) \\
\end{tabular}
\end{ruledtabular}
\end{table*}

\section{Results and discussion}

In this work we performed calculations of energy levels of the core-excited $1s2l2l'$ states of
ions along the Li isoelectronic sequence from argon ($Z = 18$) through uranium ($Z = 92$). The
corresponding energies are listed in Table~\ref{tab:en}. For each state, we present the Auger
energy $E_{\rm Auger}$ (i.e., the energy relative to the $1s^2$ state of the corresponding He-like
ion) and the energy $E$ relative to the ground $1s^22s$ state. In addition, in the last column of
the table, we present results for the fine-structure intervals $E_{\rm fs}$. The energy values
listed in the table are given with one or two uncertainties. When two uncertainties are provided,
the second one is due to the error of the nuclear charge radius, whereas the first one is the
estimate of the theoretical error. When only one uncertainty is given, it is the theoretical error,
and the one due to the nuclear radius is negligible.

Table \ref{tab:wav} presents our final results for wavelengths of the 22 strongest
$1s2l2l'$$\to$$1s^22l$ transitions. The transitions are labelled from ``$a$'' to ``$v$'', following
the widely used notations by Gabriel~\cite{gabriel:72}. Energy values are presented with one
uncertainty that includes the possible error introduced by the nuclear radii.

In the present work we obtain results not only for the energies of the core-excited states but also
for the fine-structure (fs) intervals. In most cases the fs intervals are predicted with a better
accuracy than the individual energy levels. The reason is that the QED effects for the fs intervals
are rather small, so that the QED uncertainty is almost negligible. The main uncertainty thus comes
from the DCB energies. Since we do not observe a clear correlation between the DCB energies of
different fs sublevels in our CI method, we have to propagate their errors in the standard way.

\begin{figure*}[!htb]
\centerline{
\resizebox{\textwidth}{!}{%
  \includegraphics{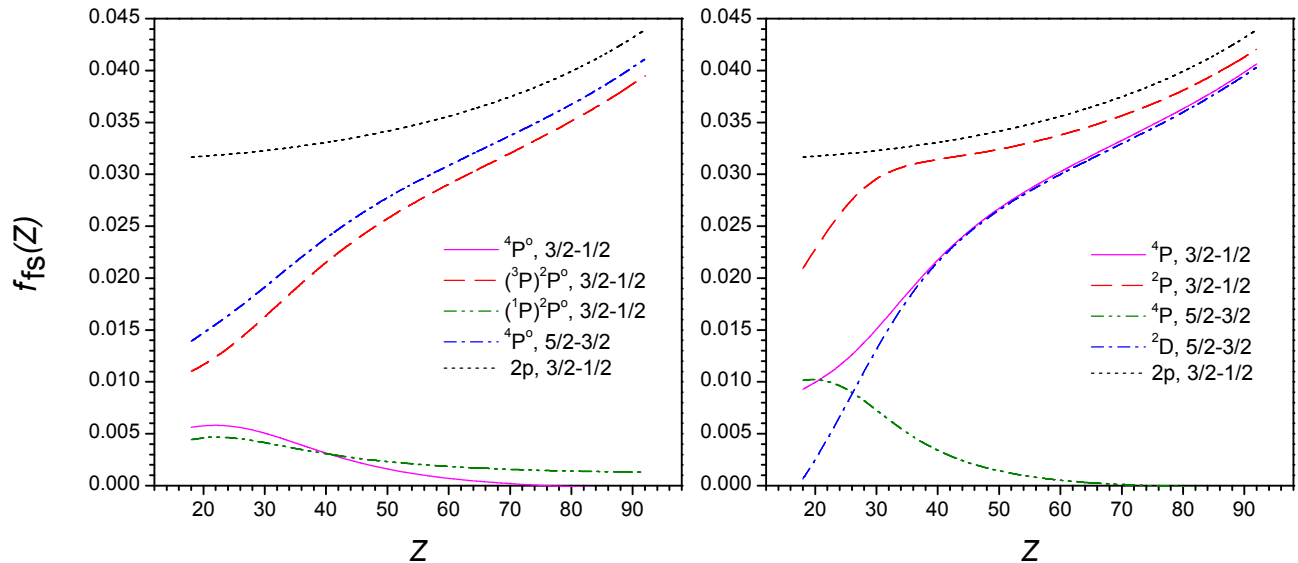}
}}
 \caption{Scaled fine-structure intervals as a function of the nuclear charge number $Z$,
 see Eq.~(\ref{eq:fs}). Left graph shows the fine-structure
 intervals of the odd ($1s2s2p$) states, whereas right graph those of the even ($1s2p^2$) states. For comparison,
 we present also the $2p_{3/2,1/2}$ fs interval of H-like ions (dotted line).
 \label{fig:1}}
\end{figure*}

In Fig.~\ref{fig:1} we present the fs intervals of the core-excited levels as a function of $Z$.
The scaled function $f_{{\rm fs}}$ is plotted, with the leading $Z$ dependence factored out,
\begin{align} \label{eq:fs}
f_{{\rm fs}}(Z,JJ') = \frac{E(J)-E(J')}{mc^2\,(Z\alpha)^4}\,.
\end{align}
For comparison, we also plot the hydrogenic $2p_{3/2,1/2}$ fs interval (dotted line). We observe
that in the low-$Z$ region individual fs intervals demonstrate very different $Z$ behaviour. It is
interesting that for the $^2D$ level, the fs interval changes its sign at around $Z = 18$ and
becomes negative for lower $Z$ \cite{yerokhin:17:lilike}.

We find that in the high-$Z$ region, the fs intervals behave in two ways, approaching either the
hydrogenic fs interval or zero. This could be readily anticipated by examining the $jj$-coupling
limit of these levels. For example, the dominant $jj$-coupling configuration of the $^4P_{3/2}^o$
level for high $Z$ is $[(1s2s)_12p_{1/2}]_{3/2}$, whereas for the $^4P_{1/2}^o$ level it is
$[(1s2s)_12p_{1/2}]_{1/2}$. We can thus expect the $^4P_{3/2,1/2}^o$ interval to become small for
high $Z$. On the contrary, the dominant $jj$-coupling configuration of the $^4P_{5/2}^o$ level for
high $Z$ is $[(1s2s)_12p_{3/2}]_{5/2}$, so it is natural that the $^4P_{5/2,3/2}^o$ interval
approaches the hydrogenic $2p_{3/2,1/2}$ fs interval for high $Z$.

We now turn to comparing our results with previous calculations and best experimental results for
several selected ions. Table~\ref{tab:Ar} shows our predictions for wavelengths of the $K\alpha$
transitions in Li-like argon, in comparison with the two best measurements
\cite{tarbutt:01,beiersdorfer:02}, the recent theoretical values obtained within the $1/Z$
expansion approach by the MZ code \cite{goryaev:17}, and the recommended NIST values \cite{nist}.
We observe a very good agreement of our theory with the experimental results, our predictions being
5-10 times more accurate than the currently available experimental data. The MZ results do not bear
any uncertainties but their maximal deviation from our values of $2\times 10^{-4}{\rm \AA}$ is
consistent with the author's accuracy expectations. The agreement of our results with the NIST
recommended values is very good, our results being more accurate by an order of magnitude.

A similar comparison for Li-like iron is presented in Table~\ref{tab:Fe}. In this case,
high-precision experimental results of Ref.~\cite{rudolph:13} are available for the $q$, $r$, $t$,
and $u$ lines. Their accuracy of $2\times 10^{-5}{\rm \AA}$ approaches the estimated precision of
our theoretical values, yielding currently the most stringent test of our calculations. The
agreement between our theory and the experimental values is very good for the $q$ and $r$ lines and
somewhat marginal (within $1.2\,\sigma$) for the $t$ and $u$ lines. Similarly to the argon case,
deviations between our results and the MZ values are consistent with the expected uncertainty of
the MZ data of $2\times 10^{-4}{\rm \AA}$. Agreement with the NIST data is again good, but it might
be noted that their uncertainty estimate of $10\times 10^{-4}{\rm \AA}$ for many levels turns out
to be overly conservative and could be decreased by a factor of 5.

In the high-$Z$ region, benchmark experimental results were obtained for Li-like mercury in
Refs.~\cite{gonzales:05:prl,gonzales:06}. In the complementary theoretical study \cite{harman:06},
dedicated calculations were performed that included estimations of theoretical uncertainty. The
corresponding comparison is presented in Table~\ref{tab:Hg}. We find agreement with both the
theoretical and the experimental results, our predictions being more precise than any of them.

Finally, Table~\ref{tab:U} compares our results for Li-like uranium with a recent calculation by
Lyashchenko and Andreev \cite{lyashchenko:16}. We observe a systematical shift of energies ranging
from 6~eV to 13~eV. The main reason for this is a $7$~eV difference in the total binding energy of
the $1s^22s$ state, which is in our case $294\,222.8\,(8)$~eV. We note that our results for the
$1s^22s$ energies are cross-checked by comparing two different methods (see
Table~\ref{tab:qedcompare}), so the deviation is most likely to be due to insufficiently accurate
treatment of QED screening effects in Ref.~\cite{lyashchenko:16}.

In summary, we have performed relativistic calculations of the energy levels and the fine-structure
intervals of the $1s2l2l'$ core-excited states of Li-like ions. The relativistic
Dirac-Coulomb-Breit energies have been obtained by the configuration-interaction method adapted for
the treatment of autoionizing core-excited states. The relativistic energies have been supplemented
with the QED energy shifts, obtained by the model QED operator approach. Relativistic recoil and
the frequency dependence of the Breit interaction has been taken into account. The theoretical
energies are supplied with uncertainty estimates. The $1s2l2l'$--$1s^22l$ transition energies are
obtained by combining the present results for the $1s2l2l'$ states with energies of the $1s^22l$
states computed by using tabulated literature data for individual QED effects. The results obtained
are in good agreement with previous theoretical and experimental data but are significantly more
accurate.

\section*{Acknowledgement}

V.A.Y. acknowledges support by the Ministry of Education and Science of the Russian Federation
Grant No.~3.5397.2017/6.7.

\begin{table*}
\caption{Comparison of theoretical and experimental wavelengths (in \AA) for transitions from the core-excited $1s2l2l'$ states
  of Li-like argon, $Z = 18$.
\label{tab:Ar}
}
\begin{ruledtabular}
\begin{tabular}{lllcccc}
\multicolumn{1}{l}{Key} & \multicolumn{1}{c}{Transition} & \multicolumn{1}{c}{Present work}
                          & \multicolumn{1}{c}{MZ code \cite{goryaev:17}}
                          & \multicolumn{1}{c}{NIST \cite{nist,saloman:10}}
                          & \multicolumn{1}{c}{Experiment \cite{tarbutt:01}}
                                                                                     & \multicolumn{1}{c}{Experiment \cite{beiersdorfer:02}}\\
\hline\\[-5pt]
$a$ & $1s2p^2(^3\!P)\,^2\!P_{3/2} \to 1s^22p\,^2\!P^o_{3/2}$   & 3.985 722\,(15) &   3.985 8   & 3.985 7\,(2)       &  3.985 67\,(15) \\
$b$ & $1s2p^2(^3\!P)\,^2\!P_{3/2} \to 1s^22p\,^2\!P^o_{1/2}$   & 3.981 663\,(15) &   3.981 8   & 3.981 7\,(2)       & \\
$c$ & $1s2p^2(^3\!P)\,^2\!P_{1/2} \to 1s^22p\,^2\!P^o_{3/2}$   & 3.989 804\,(15) &   3.989 9   & 3.989 8\,(2)       & \\
$d$ & $1s2p^2(^3\!P)\,^2\!P_{1/2} \to 1s^22p\,^2\!P^o_{1/2}$   & 3.985 738\,(15) &   3.985 8   & 3.985 7\,(2)       & \\
$e$ & $1s2p^2(^3\!P)\,^4\!P_{5/2} \to 1s^22p\,^2\!P^o_{3/2}$   & 4.012 753\,(15) &   4.012 6   & 4.012 5\,(3)       & \\
$f$ & $1s2p^2(^3\!P)\,^4\!P_{3/2} \to 1s^22p\,^2\!P^o_{3/2}$   & 4.014 759\,(15) &   4.014 6   & 4.014 5\,(2)       & \\
\end{tabular}
\end{ruledtabular}
\end{table*}

\begin{table*}
\caption{Comparison of theoretical and experimental wavelengths (in \AA) for transitions from the core-excited $1s2l2l'$  states
 of Li-like iron, $Z = 26$.
\label{tab:Fe}
}
\begin{ruledtabular}
\begin{tabular}{llllllll}
\multicolumn{1}{l}{Key} & \multicolumn{1}{c}{Transition} & \multicolumn{1}{c}{Present work}
                          & \multicolumn{1}{c}{MZ code \cite{goryaev:17}}
                          & \multicolumn{1}{c}{NIST \cite{nist,shirai:00}}
                          & \multicolumn{1}{c}{Experiment  \cite{beiersdorfer:93:aj}}
                          & \multicolumn{1}{c}{Experiment  \cite{rudolph:13}}          \\
\hline\\[-5pt]
$a$ & $1s2p^2(^3\!P)\,^2\!P_{3/2} \to 1s^22p_{3/2}\,^2\!P^o_{3/2}$  & 1.862 151 4\,(78) & 1.862 1 & 1.862 2\,(1)  & 1.862 07\,(30)             \\
$b$ & $1s2p^2(^3\!P)\,^2\!P_{3/2} \to 1s^22p_{1/2}\,^2\!P^o_{1/2}$  & 1.857 696 9\,(77) & 1.857 8 & 1.857 8\,(10) &                         \\
$c$ & $1s2p^2(^3\!P)\,^2\!P_{1/2} \to 1s^22p_{3/2}\,^2\!P^o_{3/2}$  & 1.867 270 7\,(77) & 1.867 1 & 1.867 2\,(10) &                         \\
$d$ & $1s2p^2(^3\!P)\,^2\!P_{1/2} \to 1s^22p_{1/2}\,^2\!P^o_{1/2}$  & 1.862 791 7\,(77) & 1.862 7 & 1.862 8\,(10) &                         \\
$e$ & $1s2p^2(^3\!P)\,^4\!P_{5/2} \to 1s^22p_{3/2}\,^2\!P^o_{3/2}$  & 1.872 785 1\,(77) & 1.872 6 & 1.872 8\,(10) &                         \\
$f$ & $1s2p^2(^3\!P)\,^4\!P_{3/2} \to 1s^22p_{3/2}\,^2\!P^o_{3/2}$  & 1.874 468 8\,(77) & 1.874 3 & 1.874 4\,(10) &                         \\
\end{tabular}
\end{ruledtabular}
\end{table*}

\begin{table*}
\caption{Comparision of theoretical and experimental Auger energies of the
$1s2l2l'$ states in Li-like mercury, $Z = 80$, in eV.
\label{tab:Hg}
}
\begin{ruledtabular}
\begin{tabular}{llldccc}
\multicolumn{2}{l}{ Term} & \multicolumn{1}{l}{$J$} & \multicolumn{1}{c}{Present work}
 & \multicolumn{1}{c}{MCDF \cite{harman:06}}
 & \multicolumn{1}{c}{CI-DFS \cite{harman:06}}
 & \multicolumn{1}{c}{Experiment \cite{gonzales:06}}
\\
\hline
\\[-5pt]
$1s2s^2$                   & $^2\!S$   & 1/2   &   46361.84\,(58) &  46362\,(3) & 46361\,(3) & 46358\,(4) \\
$1s       2s2p(^3\!P^o)$   & $^2\!P^o$ & 1/2   &   46615.10\,(53) &  46614\,(5) & 46614\,(4) & 46611\,(6) \\
$1s       2p^2(^3\!P)$     & $^4\!P$   & 3/2   &   48838.41\,(49) &  48842\,(5) & 48843\,(4) & 48844\,(6) \\
$1s       2s2p(^1\!P^o)$   & $^2\!P^o$ & 3/2   &   48843.62\,(54) &  48842\,(5) & 48843\,(4) & 48845\,(5) \\
$1s       2p^2(^1\!D)$     & $^2\!D$   & 3/2   &   48926.04\,(49) &  48926\,(3) & 48926\,(2) & 48918\,(9) \\
$1s       2p^2(^1\!D)$     & $^2\!D$   & 5/2   &   51062.48\,(49) &  51065\,(3) & 51064\,(2) & 51064\,(6) \\
 \end{tabular}
\end{ruledtabular}
\end{table*}

\begin{table}
\caption{Theoretical energy levels of the $1s2l2l'$ states relative to the ground state
in Li-like uranium, $Z = 92$, in eV.
\label{tab:U}
}
\begin{ruledtabular}
\begin{tabular}{llldrr}
\multicolumn{2}{l}{ Term} & \multicolumn{1}{l}{$J$} & \multicolumn{1}{c}{This work}
 & \multicolumn{1}{c}{Ref.~\cite{lyashchenko:16}$^a$}
 & \multicolumn{1}{c}{Diff.}\\
\\
\hline
\\[-5pt]
$1s2s^2$                   & $^2\!S$   & 1/2       &  95892.1\,(10)      &  95884&8     \\
$1s       2s2p(^3\!P^o)$   & $^4\!P^o$ & 1/2       &  95970.16\,(98)     &  95960&10     \\
                           &           & 3/2       &  95935.95\,(98)     &  95927&9     \\
                           &           & 5/2       & 100205.45\,(98)     &               \\
$1s       2s2p(^3\!P^o)$   & $^2\!P^o$ & 1/2       &  96225.65\,(98)     &  96214&12     \\
                           &           & 3/2       & 100325.44\,(98)     & 100315&10     \\
$1s       2p^2(^3\!P)$     & $^4\!P$   & 1/2       &  96274.93\,(94)     &  96262&13      \\
                           &           & 3/2       & 100492.73\,(94)     & 100484&9      \\
                           &           & 5/2       & 100473.66\,(94)     & 100464&10      \\
$1s       2s2p(^1\!P^o)$   & $^2\!P^o$ & 1/2       & 100400.92\,(98)     & 100394&7      \\
                           &           & 3/2       & 100536.26\,(98)     & 100530&6      \\
$1s       2p^2(^1\!D)$     & $^2\!D$   & 3/2       & 100620.67\,(94)     & 100610&11     \\
                           &           & 5/2       & 104807.93\,(94)     & 104802&6     \\
$1s       2p^2(^3\!P)$     & $^2\!P$   & 1/2       & 100530.86\,(94)     & 100521&10     \\
                           &           & 3/2       & 104899.19\,(94)     & 104892&7     \\
$1s       2p^2(^1\!S)$     & $^2\!S$   & 1/2       & 104937.25\,(96)     & 104930&7      \\
 \end{tabular}
\end{ruledtabular}
$^a$ using the energy of the ground state of $E = -294216$~keV  \cite{andreev:priv}.
\end{table}

\newpage
\begin{longtable*}{llcddd}
\caption{Energy levels of the $1s2l2l'$ states of Li-like ions, in Rydbergs. The notations are as
follows. $E_{\rm Auger}$ (the Auger energy) is the energy relative to the energy of the $1s^2$
state of the corresponding He-like ion, $E$ is the energy relative to the energy of the $1s^2 2s$
state. $E_{\rm fs}$ is the fine-structure splitting, representing the difference of the energy of
the current state and that of the previous line. $R$ denotes the nuclear charge root-mean-square
(rms) radius, whereas $M/m$ is the ratio of the nuclear mass to the electron mass. If the energy is
given with two uncertainties, the first one is the estimate of the theoretical error, whereas the
second one is due to the error of the rms radius. In the cases where only one error is given, the
second uncertainty is negligible.
 \label{tab:en}}
\\ \colrule\hline
\multicolumn{2}{c}{Term}  &
       \multicolumn{1}{c}{$J$}  &
                 \multicolumn{1}{c}{$E_{\rm Auger}$}  &
                             \multicolumn{1}{c}{$E$}  &
                                           \multicolumn{1}{c}{$E_{\rm fs}$}  \\
\hline\\            %
\multicolumn{5}{l}{
  $Z = 18,\ \mbox{\rm $^{40}$Ar},\  R = 3.4274\,(26)\ \mbox{\rm fm}, \  M/m = 72829.00$}
 \\[2pt]
$1s2s^2$                   & $^2\!S$   & 1/2       &      158.8203\,(17) &      226.3206\,(17) &                     \\
$1s2s2p(^3\!P^o)$          & $^4\!P^o$ & 1/2       &     159.40257\,(92) &     226.90283\,(92) &                     \\
                           &           & 3/2       &     159.46519\,(92) &     226.96545\,(92) &      0.062621\,(25) \\
                           &           & 5/2       &     159.62095\,(93) &     227.12121\,(93) &      0.155755\,(36) \\
\end{longtable*}

%
%
%
\begingroup
\begin{ruledtabular}
\begin{longtable*}{cllllllllllllll}
\caption{Theoretical wavelengths of the $1s 2l2l'$$\to$$1s^2 2l$ transitions in
  lithlium-like ions, in \AA. Transitions are labelled as by Gabriel \cite{gabriel:72}.
\label{tab:wav}}\\
\colrule\hline\\[-6pt]
 Key & \multicolumn{1}{c}{Transition} & \multicolumn{1}{c}{$Z = 18$} & \multicolumn{1}{c}{$Z = 19$} & \multicolumn{1}{c}{$Z = 20$} & \multicolumn{1}{c}{$Z = 21$} & \multicolumn{1}{c}{$Z = 22$} & \multicolumn{1}{c}{$Z = 23$}  \\
\\ \hline\\[-6pt]
$a$ & $1s2p^2(^3\!P)\,^2\!P_{3/2} \to 1s^22p_{3/2}\,^2\!P^o_{3/2}$   & 3.985722\,(15) & 3.562886\,(14) & 3.203646\,(12) & 2.895852\,(11) & 2.630133\,(10) & 2.3991506\,(97) \\
$b$ & $1s2p^2(^3\!P)\,^2\!P_{3/2} \to 1s^22p_{1/2}\,^2\!P^o_{1/2}$   & 3.981663\,(15) & 3.558761\,(14) & 3.199459\,(12) & 2.891610\,(11) & 2.625841\,(10) & 2.3948129\,(96) \\
$c$ & $1s2p^2(^3\!P)\,^2\!P_{1/2} \to 1s^22p_{3/2}\,^2\!P^o_{3/2}$   & 3.989804\,(15) & 3.567115\,(14) & 3.208020\,(12) & 2.900371\,(11) & 2.634790\,(10) & 2.4039384\,(96) \\
\hline\hline
\end{longtable*}
\end{ruledtabular}
\endgroup


%
%
%
\widetext

\newpage

\centerline{\bf \large Supplementary material for } \vspace*{0.25cm}

\centerline{\bf \large Energy levels of core-excited $\bm{1s2l2l'}$ states in lithium-like ions:
argon to uranium} \vspace*{0.25cm}

\centerline{\bf \large by V. A. Yerokhin and A. Surzhykov}

\vspace*{1cm}

In this Supplementary Material we present results for energies of the $1s^22s$, $1s^22p_{1/2}$, and
$1s^22p_{3/2}$ states of Li-like ions, which are compiled from different {\em ab-initio} QED
calculations available in the literature. In order to obtain the total binding energy of an
$1s^22l$ state, we added together the ionization energy of the hydrogenic $1s$ state tabulated in
Ref.~\cite{yerokhin:15:Hlike}, the ionization energy of the $1s^2$ state calculated in
Ref.~\cite{artemyev:05:pra}, and the ionization energy of the $1s^22l$ state as calculated in
Refs.~\cite{yerokhin:07:lilike,kozhedub:10}. The tabulated $1s^2$ results from
Ref.~\cite{artemyev:05:pra} were updated by replacing the one-electron two-loop QED corrections
with the more recent values from Ref.~\cite{yerokhin:15:Hlike}. The ionization energy of the
$1s^22l$ states was not presented explicitly in Refs.~\cite{yerokhin:07:lilike,kozhedub:10} and had
to be compiled from individual corrections listed therein. Specifically, we add together the
electron-correlation corrections tabulated in Ref.~\cite{yerokhin:07:lilike}, the QED screening
corrections listed in Ref.~\cite{kozhedub:10}, the few-body nuclear recoil calculated in the
present work, and the one-electron corrections from Ref.~\cite{yerokhin:15:Hlike}. The few-body
relativistic recoil correction has been calculated in Ref.~\cite{kozhedub:10} but not properly
tabulated, so we had to recalculate it. The resulting energy values for ions along the lithium
isoelectronic sequence from $Z = 18$ to $Z = 92$ are listed in Table~\ref{tab:1}.

Table~\ref{tab:2} presents a comparison of our compiled energies with the best theoretical results
available in the literature. Comparing with the results by Sapirstein and Cheng
\cite{sapirstein:11}, one should keep in mind that in that work (i) the authors used approximate
values of the nuclear charge radii from Ref.~\cite{johnson:85} for all ions except for thorium and
uranium and (ii) their uncertainties are purely theoretical and do not include possible shifts due
to errors of nuclear charge radii.

\begin{longtable}{ld}
\caption{Energy levels of the $1s^22l$ states of Li-like ions, in eV (1 eV =
$13.605\,693\,009\,(84)$~Ry \cite{mohr:16:codata}). For the $1s^22s$ state, the total binding
energy is listed; for the $1s^22p_j$ states, the energy relative to the $1s^22s$ state is
presented. If the energy value is given with two uncertainties, the first one is the estimate of
the theoretical error, whereas the second one comes from the uncertainty of the root-mean-square
nuclear charge radius $R$. In the cases where only one uncertainty is given, the second ones is
negligible. \label{tab:1} }
\\ \colrule\hline\\[-5pt]
\multicolumn{2}{l}{
  $Z = 18,\ \mbox{\rm $^{40}$Ar},\  R = 3.4274\,(26)\ \mbox{\rm fm}, \  M/m = 72829.00$}
 \\[2pt]
$1s^22s$             &    -9465.27711\,(52)(3) \\
$1s^22p_{1/2}$       &          31.86714\,(67) \\
$1s^22p_{3/2}$       &          35.03759\,(73) \\[5pt]
\multicolumn{2}{l}{
  $Z = 19,\ \mbox{\rm $^{39}$K },\  R = 3.4349\,(19)\ \mbox{\rm fm}, \  M/m = 71007.52$}
 \\[2pt]
$1s^22s$             &   -10579.41239\,(54)(3) \\
\end{longtable}

\begin{table*}
\caption{Comparison of different theoretical results for the $2p_{1/2}$-$2s$ and $2p_{3/2}$-$2s$ transition energies of Li-like
ions, in eV.
\label{tab:2}
}
\begin{ruledtabular}
\begin{tabular}{ldddd}
\multicolumn{1}{l}{$Z$}  & \multicolumn{1}{c}{Present}
                          & \multicolumn{1}{c}{Sapirstein and Cheng \cite{sapirstein:11}}
                                       & \multicolumn{1}{c}{Kozhedub {\em et al.} \cite{kozhedub:10}}
                                       & \multicolumn{1}{c}{Yerokhin {\em et al.} \cite{yerokhin:07:lilike}}\\
\hline\\[-5pt]
\multicolumn{4}{l}{$2p_{1/2}-2s$:}\\
 18                                  &      31.86714\,(67) & 31.868\,(1)   & 31.8673\,(5) \\
 20                                  &      35.96119\,(73) & 35.962\,(1)   & 35.9612\,(6) \\
 21                                  &      38.02916\,(79) & 38.031\,(1)   & 38.0289\,(7) \\
 26                                  &       48.5993\,(11) & 48.599\,(1)   & 48.5991\,(9)  & 48.6000\,(28) \\
 28                                  &       52.9510\,(13) & 52.951\,(1)   & 52.9504\,(10) & 52.9517\,(30) \\
\end{tabular}
\end{ruledtabular}
\end{table*}

\end{document}